\documentclass[10pt,conference]{IEEEtran}
\usepackage{epsfig,rotating,setspace,latexsym,amsmath,epsf,amssymb,bm,theorem}
\usepackage{cite}
\usepackage{graphicx,authblk}
\usepackage[usenames,dvipsnames,svgnames,table]{xcolor}

\newtheorem{theorem}{Theorem}
\newtheorem{lemma}{Lemma}
\newenvironment{Proof}[1]{\medskip\par\noindent
{\bf Proof:\,}\,#1}{{\mbox{\,$\blacksquare$}\par}}

\def\naive{na\"{\i}ve~}

\title{Binary Energy Harvesting Channel with \\ Finite Energy Storage\thanks{This work was supported by NSF Grants CNS 09-64364 and CNS 09-64632.}}

\author[1]{Kaya Tutuncuoglu}
\author[2]{Omur Ozel}
\author[1]{Aylin Yener}
\author[2]{Sennur Ulukus}
\affil[1]{\normalsize Department of Electrical Engineering, Pennsylvania State University, University Park, PA 16802}
\affil[2]{\normalsize Department of Electrical and Computer Engineering, University of Maryland, College Park, MD 20742}

\begin{document}

\IEEEoverridecommandlockouts
\maketitle

\begin{abstract}
We consider the capacity of an energy harvesting communication channel with a finite-sized battery. As an abstraction of this problem, we consider a system where energy arrives at the encoder in multiples of a fixed quantity, and the physical layer is modeled accordingly as a finite discrete alphabet channel based on this fixed quantity. Further, for tractability, we consider the case of binary energy arrivals into a unit-capacity battery over a noiseless binary channel. Viewing the available energy as state, this is a state-dependent channel with causal state information available only at the transmitter. Further, the state is correlated over time and the channel inputs modify the future states. We show that this channel is equivalent to an additive geometric-noise timing channel with causal information of the noise available at the transmitter. We provide a single-letter capacity expression involving an auxiliary random variable, and evaluate this expression with certain auxiliary random variable selection, which resembles noise concentration and lattice-type coding in the timing channel. We evaluate the achievable rates by the proposed auxiliary selection and extend our results to noiseless ternary channels.
\end{abstract}

\section {Introduction}

\begin{figure}[t]
\centerline{\includegraphics[width=0.85\linewidth]{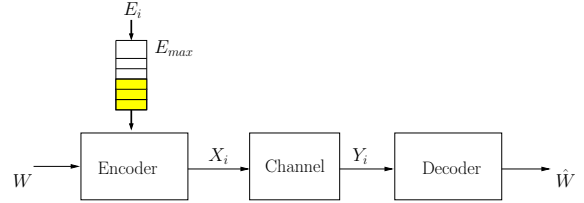}}
\caption{Energy harvesting communication system with a finite-sized battery.}
\label{system-model}
\vspace*{-0.4cm}
\end{figure}

We consider an energy harvesting communication system, where energy needed for communication is harvested by the transmitter during the course of communication; see Fig.~\ref{system-model}. We study the capacity of such communication channels. Capacity of such channels has been identified in previous work for two extreme cases: When the battery-size is unlimited, \cite{ozel2012achieving} showed that the capacity is equal to the capacity of the same system with an average power constraint equal to the average recharge rate; an unlimited-sized battery averages out the fluctuations in energy arrivals, and reduces energy causality constraints at all channel uses into a single average power constraint. At the other extreme, when the battery-size is zero, the system becomes a stochastic amplitude-constrained channel. Reference \cite{ozel2011awgn} has found the capacity of this system for an AWGN channel with independent and identically distributed (i.i.d.) energy arrivals, by modeling it as a state-dependent channel with causal state information available at the transmitter, and showed that capacity is achieved by using a Shannon strategy \cite{shannon1958channels}; in particular, the capacity achieving input distribution is discrete as in the case of \cite{Smith71}. The capacity for the case of a finite-sized battery is unknown, and is the problem considered in this paper.

This problem has the following characteristics: The battery can be viewed as an energy queue where energy arrives as a stochastic process over time; for tractability, we assume an i.i.d. energy arrival process. The codebook used to transmit messages acts as a server to this energy queue, and determines its dynamics. The energy available at the energy queue is the state, which determines the set of feasible symbols that can be transmitted. In particular, at each channel use, the transmitted symbols are instantaneously amplitude-constrained to the (square root of the) available energy in the battery. This state is naturally causally known to the transmitter, but unknown to the receiver. This state is correlated over time, even when energy arrivals are i.i.d. In addition, transmitter's own actions further affect the future of the state.

From \cite{Smith71} and \cite{ozel2011awgn}, we know that, when the channel inputs are constant amplitude-constrained, or i.i.d. stochastic amplitude-constrained, over a Gaussian channel, the optimum input distributions are discrete. However, these discrete mass points are arbitrary real numbers, and it is hard to track the dynamics of the energy queue, if it is served with codebooks generated by arbitrary real mass points. For a tractable abstraction of the system, we model energy arrivals as multiples of a fixed quantity, and correspondingly, consider a physical layer which has a discrete alphabet based on this fixed quantity. For further analytical tractability, we assume that the physical layer is a noiseless binary channel, energy arrivals are binary, and the battery is unit-sized. This abstraction is reminiscent of the one in \cite{popovski2012interactive}. We will see that, even in this simple model, unavailability of the battery state to the receiver, memory of the state in time, and the fact that the state evolves based on the previous channel inputs, render the problem challenging. The fact that channel inputs affect future states is reminiscent of action dependent channels in \cite{weissman2010capacity}. While Shannon strategy, which is optimal in the zero-battery case in \cite{ozel2011awgn}, yields achievable rates for the finite-battery case, the transmitter may utilize the memory in the battery state to achieve higher rates.

We first present the system model which is a state-dependent channel, and derive achievable rates based on certain Shannon strategies. We next show that our system may equivalently be modeled as a timing channel \cite{anantharam1996bits}, where information is transmitted by timings between 1s, as opposed to the actual places of 1s and 0s. This converts our problem into a timing channel with additive geometric noise (service time), where the service time is causally known to the transmitter. We combine \cite{anantharam1996bits} and \cite{shannon1958channels} to find a single-letter capacity expression for the capacity of this equivalent channel. Shannon strategy \cite{shannon1958channels} is optimal here, because the state is i.i.d. in time. The capacity expression involves an auxiliary random variable, and its optimization is difficult. For this reason, we determine an achievable rate based on a certain selection of this auxiliary random variable. This selection resembles the concentration idea in \cite{willems2000signaling}, and may be interpreted as a lattice-type coding for the timing channel. We evaluate the achievable rate of the proposed selection and compare it with rates achieved by Shannon strategies in the original channel. Finally, we extend our results to the case of a noiseless ternary channel by means of its equivalence to a timing channel with extra information sent inside the symbols in addition to the timings of the symbols.

\section{The Channel Model}
\label{sect_model}

Consider a noiseless binary channel with an energy harvesting transmitter. The battery in the transmitter can store at most one unit of energy. The channel input symbols, namely $0$ and $1$, have zero and one unit energy cost, respectively. When channel input $X_i$ is transmitted in the $i$th channel use, the receiver gets $Y_i=X_i$. At each channel use, the transmitter both harvests energy and also transmits a symbol. The order of harvesting and transmission in a channel use is as follows: At each channel use, $S_i$ denotes the energy available in the battery. The transmitter observes the available battery energy $S_i$ and transmits a symbol $X_i$. This symbol is constrained by the battery energy: $X_i\in\{0,1\}$ when $S_i=1$ and $X_i=0$ when $S_i=0$. After sending the symbol, the transmitter harvests energy. Energy arrivals (harvesting) is modeled as an i.i.d. Bernoulli process with $E_i\in\{0,1\}$ and $\mbox{Pr}[E_i=1]=q$. Incoming energy $E_i$ is first stored in the battery, if there is space, before it is used for transmission. Since the battery has one-unit energy storage capacity, energies may overflow and get wasted. The battery state is updated as
\begin{equation}
\label{eqn_model_update}
S_{i+1}=\min \{ S_i - X_i + E_i , 1\}.
\end{equation}

We note that $S_i$ is a state for this channel that is causally available to the transmitter but not to the receiver. Moreover, perfect channel output feedback is available to the transmitter as the channel is noiseless. Note that even though the channel is noiseless, uncertainty of the battery energy at the transmitter side makes it challenging for the receiver to decode the messages of the transmitter. A crucial property of this channel model is the fact that the state of this channel $S_i$ has memory and the evolution of the state process is affected by the channel input $X_i$. Even though the channel state is not i.i.d. in time, Shannon strategy \cite{shannon1958channels} provides some achievable rates, which we next specify for our channel model.

\subsection{Achievable Rates with Shannon Strategies}
\label{sub_mock}

Consider the Shannon strategy $(0,0)$ and $(0,1)$ as in \cite{shannon1958channels} where the first and second entries denote the actual channel input when battery state $S$ is $0$ and $1$, respectively. We represent $(0,0),(0,1)$ as $U=0,1$ for simplicity, and let $U$ be i.i.d. with $\mbox{Pr}[U=1]=p$. The state transition probabilities are $\mbox{Pr}[S_{i+1}=1|S_{i}=0]=q$ and $\mbox{Pr}[S_{i+1}=0|S_{i}=1]=p(1-q)$, yielding the stationary probability $\mbox{Pr}[S=1]=\frac{q}{p+q-pq}$. Note that the battery state is ergodic and the receiver can use the stationary probability of the battery state for joint typicality decoding. In this case, the channel takes the form  
\begin{align}
p(y|u)=\mbox{Pr}[S=1]\delta(y-u) + (1-\mbox{Pr}[S=1])\delta(y)
\end{align}

\noindent where $\delta(\cdot)$ is a unit impulse at $0$ and $u \in \{0,1\}$. Here, $I(U;Y)$ is an achievable rate due to \cite{shannon1958channels} and the fact that the battery energy state is an ergodic process, see also \cite{popovski2012interactive}. We refer to this as the \naive i.i.d. Shannon strategy (NIID). The best achievable rate by this strategy is
\begin{align}
R_{NIID} = \underset{p \in [0,1]}{\max}~ H \left( \frac{pq}{p+q-pq} \right)- p H \left( \frac{q}{p+q-pq} \right)
\end{align}
where $H(\cdot)$ is the binary entropy function.

The \naive approach yields a sub-optimal rate since the memory of the system is not exploited by the decoder. Indeed, the decoder can better exploit the memory of the system by using the $n$-letter joint probability $p(u^n,y^n)$. The maximum rate that can be achieved by using this scheme is given by
\begin{align}
\label{eqn_roiid}
R_{OIID}=\max_{p \in [0,1]} \lim_{n \rightarrow \infty} \frac{1}{n}I(U^n;Y^n).
\end{align}

\noindent We will call the scheme that achieves $R_{OIID}$ the optimal i.i.d. Shannon strategy (OIID). Here, the limit $\lim_{n \rightarrow \infty} \frac{1}{n}I(U^n;Y^n)$ can be calculated numerically using the method in \cite{arnold06}. Notice that when the input $u_i$ is i.i.d., the joint probability $p(y_i,u_i,s_{i+1}|s_i)$ can be expressed in the form 
\begin{align}
p(y_i,u_i,s_{i+1}|s_i)=p(y_i,s_{i+1}|u_i,s_i)p(u_i)
\end{align} 

\noindent where $p(y_i,s_{i+1}|u_i,s_i)$ is independent of $i$. This allows using the iterative method in \cite{arnold06} to calculate the achievable rate for a given input distribution, which is then optimized with respect to the i.i.d. probability distribution to yield (\ref{eqn_roiid}).

\section {Equivalent Timing Channel}
\label{sect_equivalent}

\begin{figure}[t]
\centerline{\includegraphics[width=0.8\linewidth]{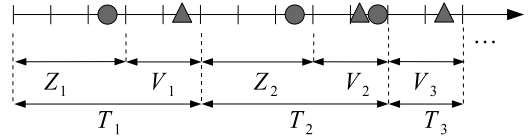}}
\caption{Graphical representation of $T_n$, $V_n$ and $Z_n$. An energy arrival of $E_i=1$ is represented with a circle and a channel input of $X_i=1$ is represented with a triangle. Note that $Z_3=0$ as a transmission of a $1$ and an energy harvesting occur in the same channel use.}
\label{fig_model}
\vspace*{-0.4cm}
\end{figure}

Since the channel input is binary, the encoding and decoding can be performed over the number of channel uses between two $1$s. Let us define $T_n \in \{1, 2, \ldots\}$ as the number of channel uses between the $n-1$st transmitted $1$ and the $n$th transmitted $1$. After a $1$ is transmitted in the $i$th channel use, in view of (\ref{eqn_model_update}), available energy in the battery drops to zero in the $i+1$st channel use unless $E_{i} = 1$ as the battery can store at most one unit of energy. The node cannot transmit another $1$ until the next energy arrival. Define the idle time $Z_n \in \{0,1,2,\ldots\}$ as the number of channel uses between the $n-1$st transmission of a $1$ and the next energy arrival after that. This representation yields the following timing channel:
\begin{equation}
T_n=V_n+Z_n
\end{equation}
where $V_n \in \{1,2,\ldots\}$ is the number of channel uses the transmitter chooses to wait to transmit a $1$ after the first energy arrival proceeding the $n-1$st transmission of a $1$. $T_n$, $V_n$ and $Z_n$ are depicted in Fig.~\ref{fig_model}, where circles represent energy arrivals and triangles represent transmissions of a $1$. Note that $Z_n$ is independent of any action of the transmitter; it is the number of i.i.d. Bernoulli trials until the first energy arrives after the $n-1$st transmission of a $1$, and is a geometric random variable with parameter $q$. As the original channel is noiseless, when the transmitter puts a $1$ to the channel, the receiver observes it and calculates $T_n$ perfectly. Therefore, the transmitter observes $Z_n$ and decides on $V_n$, but all that the receiver observes is $T_n$. Even though the physical channel is noiseless, the uncertainty in $Z_n$ creates difficulty for the receiver. In fact, in terms of a timing channel, the overall channel is an additive noisy timing channel with a geometric noise, which is known causally to the transmitter.

It is worthwhile to mention that when the channel symbols are considered as the time difference between two 1s, the resulting model becomes similar to a noiseless channel with symbols of varying durations as in Shannon's original work \cite{shannon1948}. However, in our problem, the symbol durations are affected by a random energy arrival process, while in \cite{shannon1948} the symbol durations are fixed. Therefore, the problem in \cite{shannon1948} is to \textit{pack} as many symbols as possible within a given block length, while our problem is also concerned with the randomness introduced by energy harvesting.

In the timing channel, the transmitter has the feedback of previous channel outputs $T^{n-1}$ and causal knowledge of the idle time $Z_n$ before deciding $V_n$. In fact, the transmitter observes energy arrivals $E_i$ even when the battery is full. We note that the timing channel in this section and the classical state-dependent channel in Section~\ref{sect_model} are equivalent in the sense that they have the same capacity, using the definition of capacity of a timing channel in \cite{anantharam1996bits}. This is due to the fact that the encoders and decoders of these channels have different representations of the same object, and the rate and capacity of these channels are properly defined taking into account the time cost of the codewords, i.e., the average number of channel uses needed, following \cite{anantharam1996bits}. We state this fact as a lemma. 

\begin{lemma}
\label{lem_equivalent}
The timing channel capacity with additive causally known state $C_{T}$ and the classical state-dependent channel capacity $C$ are equal, i.e., $C=C_T$.
\end{lemma}

\section{The Capacity}
\label{sect_capacity}

The capacity of a discrete memoryless channel with causal side information is \cite{shannon1958channels}, \cite[Thm. 7.2]{el2011network}
\begin{equation}
\label{eqn_capacity_timing}
	C=\underset{p(u),v(u,z)}{\max}~ I(U;T),
\end{equation}
where $T$ is the channel output. The maximization requires finding the optimal distribution for the auxiliary variable $U$ and the mapping $v(\cdot,\cdot)$ which can be taken as deterministic without losing optimality \cite[Thm. 7.2]{el2011network}. Cardinality bound on $U$ is $|\mathcal{U}| \leq \min \{(|\mathcal{V}|-1)|\mathcal{Z}|+1,|\mathcal{T}|\}$. We combine techniques from \cite{shannon1958channels} and \cite{anantharam1996bits} to prove the following theorem.

\begin{theorem}
\label{thm_capacity}
The capacity of the timing channel with additive causally known state, $C_{T}$, is:
\begin{equation}
\label{eqn_capacity1}
	C_{T}=\underset{p(u),v(u,z)}{\max}~ \frac{I(U;T)}{\mathbb{E}[T]}
\end{equation}
\end{theorem}

\begin{Proof}
Let $W \in \{1,\ldots,M\}$ denote the message, which is uniform and let $n$ denote the maximum number of channel uses on average (over the message $W$ and the energy arrivals $E_i$) to send any message $W=w$. We note that $\sum_{i=1}^{m} \mathbb{E}[T_i] \leq n$ where the expectation is over the energy arrival sequence $E_i$ and the message $W$.

Define $U_i = (W,T^{i-1})$. Note that even though $T^i$ depends on $W$, since $E_i$ is an i.i.d. random process, $Z_i$ is independent of $W$ and $T^{i-1}$ and hence of $U_i$. We have the following sequence of inequalities:
\begin{align}
\log(M) - H(W|T^m) &= H(W) - H(W|T^m)\\
&= I(W;T^{m}) \label{bir}\\
&= \sum_{i=1}^{m} I(W;T_i | T^{i-1}) \\
&\leq \sum_{i=1}^m I(W,T^{i-1};T_i) \label{iki} \\
&= \sum_{i=1}^m I(U_i;T_i) \\
&\leq \frac{n}{\sum_{i=1}^m \mathbb{E}[T_i]} \sum_{i=1}^m I(U_i;T_i) \label{uc} \\
&\leq n \sup_{U} \frac{I(U;T)}{\mathbb{E}[T]} =nC_T \label{bes}
\end{align}
where (\ref{uc}) follows from the fact that $n \geq \sum_{i=1}^m \mathbb{E}[T_i]$, and the inequality in (\ref{bes}) holds due to the fact that $U_i$ are independent of $Z_i$ and $\frac{\sum_i a_i}{\sum_i b_i} \leq \max_i \frac{a_i}{b_i}$ for $a_i,b_i >0$. Whenever the probability of error goes to zero as $m \rightarrow \infty$, $H(W|T^m) \rightarrow 0$ by Fano's inequality and therefore, $\frac{\log(M)}{n}=R \leq C_T$, which completes the converse proof.

For the direct part, we simply use an encoding scheme that does not use the available feedback information $T^{i-1}$ and the achievability of $R=\frac{I(U;T)}{\mathbb{E}[T]}$ follows from \cite{shannon1958channels}, \cite{anantharam1996bits} and \cite{el2011network}.
\end{Proof}

We note that the capacity expression in (\ref{eqn_capacity1}) is similar to the capacity of a telephone signaling channel \cite{anantharam1996bits}, augmented with Shannon strategy to utilize the causal state information at the transmitter. It is worth remarking that the transmitter, in fact, has more freedom in that, it is free to update its decision after observing $Z^i$; however, as the converse proof of Theorem~\ref{thm_capacity} shows, this does not yield a higher achievable rate.

For the timing channel, the input, output and state alphabets have infinite cardinalities. Therefore, the cardinality bound on $U$ is infinite and the solution to the maximization problem in (\ref{eqn_capacity1}) is difficult. While the capacity achieving auxiliary selection is still an open problem, we continue our capacity analysis by resorting to a family of auxiliary random variables with finite cardinality, in the next section. We will also devise an upper bound in Section~\ref{sect_ub} in closed form which will help us assess the performance of the proposed finite cardinality auxiliary random variables.

\section{Finite Cardinality Auxiliary Variables}
\label{sect_achievable}

In this section, we propose to use a finite cardinality auxiliary random variable selection parameterized by a variable $N$ to be optimized later. Let $U$ have a probability mass function (pmf) $p(u)$ over a support set $U\in\{0,1,\ldots,N-1\}$, where $N$ is the cardinality of $U$. Moreover, we fix the mapping from $U$ and $Z$ to $V$ as
\begin{equation}
\label{eqn_achievable_v}
	V=(U-Z \mbox{ mod } N) + 1
\end{equation}
The output of the timing channel is $T=V+Z$. Let $T^\prime=T-1 \mbox{ mod }N$. Note that $T^\prime = U$. We have
\begin{align}\label{xx}
\underset{p(u)}{\max}~ \frac{I(U;T)}{\mathbb{E}[V+Z]} &\leq \underset{p(u)}{\max}~ \frac{H(U)}{\mathbb{E}[V+Z]} \\ &= \underset{p(u)}{\max}~ \frac{I(U;T^\prime)}{\mathbb{E}[V+Z]} \label{eqn_achievable}
\end{align}
where (\ref{xx}) is due to the non-negativity of entropy. Therefore, for fixed $N$, this scheme achieves \begin{align}
R_{A}^{(N)} = \underset{p(u)}{\max}~ \frac{H(U)}{\mathbb{E}[V+Z]}
\end{align}
We further maximize $R_{A}^{(N)}$ over $N$ and obtain the best achievable rate by this scheme: $R_{A}=\underset{N}{\max}~ R_{A}^{(N)}$.

\begin{figure}[t]
\centerline{\includegraphics[width=0.8\linewidth]{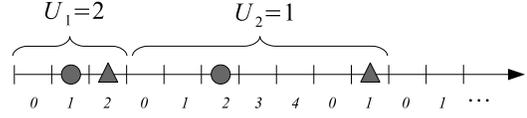}}
\caption{Coding scheme: each message symbol $U_i$ is conveyed by transmitting a $1$ at the earliest channel use possible with index equal to $U_i$. Here, $N=5$.}
\label{fig_coding}
\vspace*{-0.4cm}
\end{figure}

In the original binary energy harvesting channel, this strategy corresponds to channel uses being divided into frames of length $N$ and indexed in base $N$. The transmitter constructs a codebook with symbols $U\in\{0,1,\ldots,N-1\}$ and conveys each symbol by transmitting a $1$ at the earliest channel use indexed with this symbol as illustrated in Fig.~\ref{fig_coding} for $N=5$. This coding strategy can be interpreted as concentrating the effective state $Z$ to multiples of a {\it frame length} $N$ as in \cite{willems2000signaling}. The receiver decodes intended codeword $U$ with $U=T-1\mbox{ mod }N$, which treats $Z$ as if it is eliminated with the modulo operation. Therefore, this scheme is a variation of the concentration scheme of Willems in \cite{willems2000signaling} for the timing channel. It can also be interpreted as lattice-coding the timing via modulo operation.

\section{An Upper Bound}
\label{sect_ub}

In order to get an upper bound, we provide the channel state of the timing channel $Z$ to the receiver.
\begin{align}
\label{eqn_capacity_ob1}
	C_{UB} 	&=\underset{p(v)}{\max}~ \frac{H(V)}{\mathbb{E}[V]+\mathbb{E}[Z]}\\
\label{eqn_capacity_ob3}
	& =\underset{\mu \geq 0}{\max}~ \frac{1}{\mu+\mathbb{E}[Z]} \underset{\mathbb{E}[V] \leq \mu}{\max}~ H(V)	
\end{align}
Here, the inner problem requires finding the entropy maximizing probability distribution with an expectation constraint of $\mu$ over a discrete positive support, which is solved by a geometrically distributed $V$ with parameter $\frac{1}{\mu}$. Thus, $H(V)=\frac{H(p)}{p}$ where $H(p)$ is the binary entropy function. Substituting in (\ref{eqn_capacity_ob3}) and noting that $Z\in\{0,1,\ldots\}$ is geometrically distributed with parameter $q$, the upper bound is found as
\begin{equation}
\label{eqn_capacity_ob4}
	C_{UB} =\underset{p \geq 0}{\max}~ \frac{H(p)/p}{\frac{1}{p}+\frac{1-q}{q}} = \underset{p \geq 0}{\max}~ \frac{qH(p)}{q+p(1-q)}
\end{equation}

\section{Numerical Results}
\label{sect_numerical}

In this section, we provide simple numerical results for the achievable scheme proposed in Section~\ref{sect_achievable}, how it compares to the i.i.d. Shannon strategies in the classical state-dependent channel given in Section~\ref{sub_mock}, and the upper bound in Section~\ref{sect_ub}. We also provide comparisons with the capacities in the extreme cases of batteries with zero and unlimited energy storage (still operating over a noiseless binary channel). Next, we briefly mention the capacities under these extreme cases.

\subsection{Capacity with Zero Energy Storage}
\label{sub_ns}

Consider the model with no energy storage, that is, $X_i=1$ is allowed only if $E_i=1$, which is known causally by the transmitter. In this case, the state of the system, $E_i$, is i.i.d. The capacity of this channel is achieved by a Shannon strategy as in \cite{ozel2011awgn}, which chooses among $X_i\in\{0,1\}$ with $\mbox{Pr}[X_i=1|E_i=1]=p$ whenever $E_i=1$, and is forced to transmit $X_i=0$ whenever $E_i=0$. The capacity is:
\begin{align}
\label{eqn_nostorage}
C_{ZS} &= \underset{p}{\max}~ H(pq)-pH(q)
\end{align}
where $H(p)$ is the binary entropy function.

\subsection{Capacity with Infinite Energy Storage}
\label{sub_is}

When there is an unlimited battery, as the average energy arrival rate is $q$, the capacity is the maximum of $H(p)$ over $p \leq q$, which is achieved by using the save-and-transmit scheme in \cite{ozel2012achieving}. When evaluated, the capacity becomes:
\begin{equation}
C_{IS}=
	\begin{cases}
		H(q), & \quad q \leq \tfrac{1}{2} \\
		1, & \quad q > \tfrac{1}{2}
	\end{cases}
\end{equation}

\begin{figure}[t]
\centerline{\includegraphics[width=0.98\linewidth]{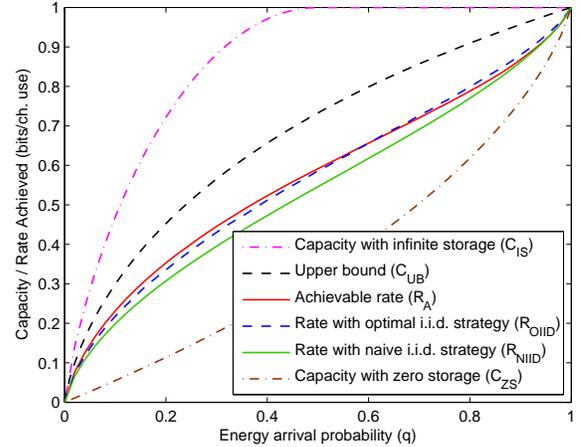}}
\caption{Upper bound, achievable rate, no battery capacity, infinite battery capacity, and rates by \naive and optimum i.i.d. Shannon strategies.}
\label{fig_plot}
\vspace*{-0.3cm}
\end{figure}

We are now ready to compare the capacities and achievable rates, which are shown in Fig.~\ref{fig_plot} as a function of the energy arrival probability $q$. We first remark that the achievable rate by the specified family of finite cardinality auxiliary variables in Section \ref{sect_achievable} is higher than the best achievable rate by the \naive i.i.d. Shannon strategy in the classical state-dependent channel where the receiver uses the stationary distribution of the battery energy. This achievable rate also outperforms the optimum i.i.d. Shannon strategy for low arrival probabilities. We also remark that in comparison to zero and infinite energy storage capacities, a battery of unit size provides a significant rate improvement.

For the achievable rate proposed in Section~\ref{sect_achievable}, a large $N$ provides a larger cardinality for $U$, sending more information with a single unit of energy. However, as $N$ increases, each symbol takes more time and more harvested energy is potentially wasted. We observe numerically that for low harvest rates the optimal $N$ decreases with increasing $q$ so that more of the scarcely available energy is utilized. This trend is reversed for high harvest rates, e.g., $q>0.7$, as wasting energy becomes less of a concern then.

\section{Extension to Noiseless Ternary Channel}
\label{sect_ternary}

In this section, we consider an extension with a third channel input $-1$, which also requires one unit of energy. Specifically, we represent the ternary output vector $Y^n$ with $T^\ell \in \{1,2,\ldots\}^\ell$, defined as the number of zeros between the $\ell-1$st and the $\ell$th nonzero output, and $S^\ell \in \{-1,1\}^\ell$, defined as the sign of the $\ell^{th}$ nonzero output. Once again, note that $Y^n$ can be perfectly reconstructed from $(T^\ell,S^\ell)$ and vice versa. With this representation, the equivalent model consists of two parallel channels, $T_n=V_n+Z_n$ and $S_n=R_n$, where $R_n$ is the sign of the $\ell$th nonzero input, conveyed noise-free to the receiver. The equivalence to the original channel can be established by an extension of Lemma~\ref{lem_equivalent}. Noting that the binary side channel and the timing channel are used equally many times, the information conveyed with each of their use is $I(U;T) + 1$ where $U$ is the Shannon strategy. This is similar to queues with information bearing packets, since each nonzero symbol can be considered as a packet bearing $1$ bit of information. Hence, as in \cite[Section IV]{anantharam1996bits}, coding over symbol energy and over symbols of the same energy are done independently and this yields the capacity.

\begin{theorem}
The capacity with ternary symbols is:
\begin{align}
\label{eqn_cap_ternary}
	C_{ter} = \underset{p(u),v(u,s)}{\max}~ \frac{I(U;T) + 1}{\mathbb{E}[T]}
\end{align}
\end{theorem}

\section{Conclusion}
\label{sect_conclusion}

In this paper, we considered an energy harvesting transmitter with a unit-capacity battery communicating over a noiseless binary channel. Classical view of this channel is a state-dependent channel with causal state information at the transmitter where the evolution of the state is affected by the channel input. This coupling of the state and the channel input creates a challenge. We show that this channel is equivalent to a timing channel with feedback and causal state information at the transmitter, and find a single-letter capacity expression for its capacity. As evaluation of this capacity expression is difficult, we resort to finite cardinality auxiliary random variables and numerically study the achievable rates. We compare these achievable rates with rates achievable by \naive and optimum i.i.d. Shannon strategies in the original channel. Finally, we extend our results to a noiseless ternary channel where information is transmitted via timing and contents of symbols.

\bibliographystyle{IEEEtran}
\bibliography{IEEEabrv,bib_eh}

\end{document}